\documentclass[sn-mathphys-num]{sn-jnl}

\usepackage{graphicx}%
\usepackage{multirow}%
\usepackage{amsmath,amssymb,amsfonts}%
\usepackage{amsthm}%
\usepackage{mathrsfs}%
\usepackage[title]{appendix}%
\usepackage{xcolor}%
\usepackage{textcomp}%
\usepackage{manyfoot}%
\usepackage{booktabs}%
\usepackage{algorithm}%
\usepackage{algorithmicx}%
\usepackage{algpseudocode}%
\usepackage{listings}%

\theoremstyle{thmstyleone}%
\theoremstyle{thmstyletwo}%

\theoremstyle{thmstylethree}%

\begin{document}

\title[Article Title]{Electron scattering on $^4$He from coupled-cluster theory}


\author*[1]{\fnm{Joanna E.} \sur{Sobczyk}}\email{jsobczyk@uni-mainz.de}

\affil*[1]{\orgdiv{Institut f\"ur Kernphysik and PRISMA$^+$ Cluster of Excellence}, \orgname{Johannes Gutenberg-Universit\"at}, \orgaddress{\city{Mainz}, \postcode{55128}, \country{Germany}}}


\abstract{We present a coupled-cluster calculation for the electron-$^4$He scattering  in the region of the quasi-elastic peak. We show the longitudinal and transverse responses separately, and discuss results  within two distinct theoretical methods: the Lorentz integral transform and spectral functions. The comparison between them allows to investigate the role of final state interactions, two-body currents and relativistic effects.}


\maketitle

\section{Introduction}\label{sec:intro}

In recent years, there has been a resurgence of interest in lepton-nucleus scattering, particularly due to long baseline neutrino oscillation programs~\cite{Nustec, whitepaper2}. These experiments aim at measuring fundamental neutrino properties, such as mixing angles and, notably, the CP-violating phase. These are inferred from the observed oscillation patterns, the accuracy of which hinges upon the precise reconstruction of neutrino energy in scattering events. This reconstruction procedure is heavily reliant on theoretical models of lepton-nucleus interactions. 
 It is widely acknowledged that electron scattering processes, benefiting from precise datasets, offer an ideal opportunity to validate theoretical calculations~\cite{whitepaper1}. These two processes are analogous, with the neutrino-induced reactions being theoretically more sophisticated since neutrinos break parity and so the process involves both vector and axial currents.
Historically, electromagnetic probes have been recognized as excellent tools for investigating the structural properties of nuclear systems. Electron scattering experiments conducted at various energy scales provide precise data across a wide energy range. This breadth of data is crucial for neutrino experiments, as their energy range of interest, spanning from hundreds of MeV to a few GeV, encompasses various physical mechanisms,  from the quasi-elastic (QE) peak to deep inelastic scattering.
Already in the QE peak the typical energies relevant for experiments are high, so that the relativistic treatment in theoretical calculations becomes crucial.

Future oscillation experiments, DUNE~\cite{DUNE} and Hyper-K~\cite{hyperk}, use detectors built of medium-mass nuclei ($^{40}$Ar and $^{16}$O respectively), making the many-body nuclear methods an essential tool to interpret experimental data.
Motivated by this need, we initiated a program which aims at calculating neutrino-nucleus cross sections from an ab initio coupled-cluster framework~\cite{kuemmel1978,dean2004,Hagen:2010gd}. The ultimate goal of this endeavor are medium-mass nuclei of experimental interest. However, the method should be first benchmarked in the few-body sector for electromagnetic reactions for which highly precise few-body approaches exist.
In the following, we present results in the QE regime from the coupled-cluster calculations for $^4$He which is an excellent case-study, being a closed-shell nucleus. We discuss two complementary  approaches, the Lorentz integral transform and spectral functions, which are reliable at low and high momentum transfers respectively. 
Spectral functions, assuming certain approximations, allow for the relativistic calculation required in the large phase-space probed by neutrino experiments.
For moderate momenta, the comparison of two approaches reveals the role of the used approximations.

\section{Formalism}\label{sec:formalism}

Inclusive electron scattering cross section $(e,e')$ can be written in terms of two response functions,
\begin{equation}
    \frac{d^2\sigma}{d\Omega_{e'} d\omega} = \sigma_{Mott} \left( \left(\frac{q^2}{{\bf q}^2}\right)^2 R_L(\omega,|{\bf q}|) + \left(\tan^2\frac{\theta_{e'}}{2}-\frac{1}{2}\frac{q^2}{{\bf q}^2}\right) R_T(\omega,|{\bf q}|) \right),
\end{equation}
with the Mott cross section $\sigma_{Mott}$, the energy-momentum transfer $q=(\omega, {\bf q})$ and the scattering angle $\theta_{e'}$ of an outgoing electron. 
The nuclear responses $R_{L/T}$ encapsulate the behaviour of the probed nucleus. At low energies they describe excitations of low-lying states and resonances. For the momentum transfers ${\bf q}$ of the order of $\mathcal{O}(100)$ MeV  the response is dominated by a single nucleon knock-out reaction in the so-called QE regime.
Within the nonrelativistic picture, the responses correspond to the excitations of the system by the charge operator and the transverse part of the electromagnetic current:

\begin{equation}
    R_L(\omega,|{\bf q}|) = \sum_f \langle 0|\rho^\dagger({\bf q})|f\rangle \langle f| \rho({\bf q})|0\rangle \delta(E_0+\omega-E_f)\ ,
\end{equation}

\begin{equation}
    R_T(\omega,|{\bf q}|) = \sum_f \langle 0|{\bf j_T}^\dagger({\bf q})|f\rangle \langle f| {\bf j_T}({\bf q})|0\rangle \delta(E_0+\omega-E_f)\ .
\end{equation}
The one-body operators $(\rho, {\bf j})$ have a standard form. They act on the groundstate nucleus $|0\rangle$, exciting it to the final $|f\rangle$ state.
The sum over $f$ is a highly demanding computational problem since it requires techniques which give access to a continuum set of states. 
A well-established approach to solve this problem within ab initio nuclear methods, amounts to calculating an integral transform of the response instead,
\begin{equation}
	 \tilde{R}(\sigma,|{\bf q}|) =  \langle 0|\theta^\dagger K\left(\sigma-E_0,\mathcal{H} \right)  \theta |0\rangle
\end{equation}
 and inverting it afterwards. Here, $\theta$ is an operator of interest (in our case $\rho$ or ${\bf j}$), $K$ corresponds to a kernel function and $\mathcal{H}$ is the nuclear Hamiltonian.
For the few-body sector, the Lorentzian and Laplace kernels have been successfully employed~\cite{Efros:1994iq, efros2007,Carlson2015,Lovato_2016}. In last years the Lorentz integral transform (LIT) has been used in conjuncture with the many-body coupled-cluster framework, first to obtain the dipole response of medium-mass nuclei~\cite{bacca2013,bacca2014} and more recently also for the longitudinal and transverse responses~\cite{Sobczyk:2021dwm,Sobczyk:2023sxh}.
This computational procedure leads to results in which the final excited states -- albeit explicitly not known -- are treated in a consistent way, i.e. not introducing any approximation to the form of the final wave function.
However, this approach comes with some drawbacks. Firstly, it is burdened with a heavy computational cost. Secondly, 
the inversion procedure is an ill-posed problem which requires some additional techniques or smoothing assumptions about the shape of the response.

The cross section in the QE regime can be approximated, assuming that a single-nucleon knock-out process occurs and the final nucleon does not interact with the residual system. 
In this picture, called impulse approximation (IA), the nuclear information is encoded in the spectral functions (SF) which gives the energy-momentum distribution of removing a nucleon from the groundstate
\begin{align}
	S(E, \mathbf{p})&=\sum_{\alpha,\alpha'}\int_{\Psi_{A-1}}  |\langle 0|a_\alpha^\dagger  |\Psi_{A-1}\rangle \langle \Psi_{A-1}| a_{\alpha'}| 0\rangle  \nonumber \\
	&\langle \mathbf{p}|\alpha\rangle^\dagger \langle \mathbf{p}|\alpha'\rangle \delta(E+E_{f}^{A-1}-E_0)\,,
	\label{eq:SF_def}
\end{align}
where $a_\alpha$ is a one-particle removal operator with quantum numbers $\alpha$.
The nuclear responses can be then directly related to the  SF
%
\begin{equation}
R_{\mathrm{IA}}(\omega, {\bf q}) = \int d^3{\bf p} dE\ S(E, {\bf p}) | \langle {\bf p}+{\bf q} | \theta| {\bf p}\rangle|^2  \delta(E+\omega-E_{p+q})\,.
\label{eq:IAnonrel}
\end{equation}
The knocked-out nucleon at interaction vertex is treated as a free nonrelativistic particle,  leading to a drastic simplification of the response calculation, in which  the matrix element $| \langle {\bf p}+{\bf q} | \theta({\bf q})| {\bf p}\rangle|^2$ is factorized from the information of the groundstate nucleus.
This factorization allows to extend the formalism to the relativistic regime: while the spectral function is obtained using nonrelativistic nuclear techniques, it can be coupled to the vertex of interaction in which both the kinematics and the electromagnetic current can be treated in the relativistic theory
\begin{equation}
	R_{\mathrm{IA}}^{\mathrm{rel}}(\omega, {\bf q}) = \int d^3{\bf p} dE\, \frac{m}{E_p} \frac{m}{E_{p+q}} S(E, {\bf p}) | \langle p+q| \theta_{\mathrm{rel}}(q)| p\rangle|^2  \delta(E+\omega-E_{p+q})\,,
	\label{eq:IArel}
\end{equation}
with the relativistic energy $E_{p+q} = \sqrt{m^2+({\bf p}+{\bf q})^2}$ and additional $m/E$ factors which arise from the normalization of Dirac spinors. One can also include the relativistic effects only in the kinematics leading to
\begin{equation}
	R_{\mathrm{IA}}^{\mathrm{semi-rel}}(\omega, {\bf q}) = \int d^3{\bf p} dE\, \frac{m}{E_p} \frac{m}{E_{p+q}} S(E, {\bf p}) | \langle {\bf p}+{\bf q} | \theta| {\bf p}\rangle|^2   \delta(E+\omega-E_{p+q})\,.
	\label{eq:IAsemirel}
\end{equation}
It has to be noted that for the relativistic currents, which depend on the four-momentum transfer, the procedure is not fully consistent since the initial nucleon is off-shell (with the energy-momentum distributed according to the spectral function). On the other hand, the vertex of interaction assumes both initial and final nucleons to be on-shell.  This leads to some ambiguities and a violation of current conservation~\cite{Benhar:2006wy} (which can be restored by de Forest procedure~\cite{DeForest:1983ahx}).
%

\subsection{Coupled-cluster for LIT and SF}

Within the coupled-cluster method, the nuclear A-body groundstate wavefunction is constructed based on an ansatz:
\begin{equation}
	|0\rangle= e^T |\Psi_0\rangle
\end{equation} 
with the correlation operator $T=T_1+T_2+...$ and the Hartree-Fock state $|\Psi_0\rangle$. The $T_n$ operator creates n-particle n-hole excitations in respect to the reference state.  The sum of $T=\sum_i T_i$ is truncated at some level. The results presented in this work were obtained within singles and doubles approximation (CCSD), i.e. $T=T_1+T_2$. The $T$ amplitudes are obtained by solving a set of coupled equations.
The dynamics between constituent nucleons is modelled by state-of-the-art chiral Hamiltonians. 
They are derived from chiral perturbation theory and allow for an order-by-order expansion with low energy constants fit to experimental data. In our calculations, we employ NNLO$_\text{sat}$ potential with two- and three-body forces, which was adjusted to groundstate properties of some medium-mass nuclei~\cite{Ekstrom:2015rta}.
We perform the calculation of LIT and SF both within the same many-body method (and approximations within) and employing the same nuclear dynamics. This allows us to perform a meaningful comparison of both approaches to understand the role of various approximations.

The calculation of nuclear response functions using LIT is performed within a nonrelativistic formalism, both in the description of nuclear dynamics, as well as in electromagnetic currents. 
The scattering problem is formally reformulated into a bound-state problem.
The Lorentz Integral Transform, merged with the coupled-cluster method (LIT-CC),  is based on the equation-of-motion technique~\cite{stanton1993}. The inversion is performed by the least squares method~\cite{efros2007}.

The spectral function for $^4$He within the coupled-cluster formalism was obtained in Ref.~\cite{Sobczyk:2022ezo}. The calculation was performed using Gaussian Integral Transform, through an expansion into Chebyshev polynomials. In this sense the procedure resembles LIT calculation. However, rather than inverting the integral transform, the final spectral reconstruction is built as a histogram, allowing for a rigorous estimation of uncertainties~\cite{Sobczyk:2021ejs}.
As mentioned in Sec.~\ref{sec:formalism}, within IA the spectral function can be directly employed to calculate the cross section.

We note that for light systems the coupled-cluster method is prone to the center-of-mass spurious states~\cite{hagen2009a}. This problem was addressed in Ref.~\cite{Sobczyk:2020qtw} for the LIT and Ref.~\cite{Sobczyk:2022ezo} for the SF, in both cases assuming that the center-of-mass motion is approximately a Gaussian factorized from the wavefunction. We found that the spurious states are excited at low energy transfers and can be removed from the spectrum. For the SF, the effect can be mitigated by a transformation of the momentum distribution.

\section{Results}\label{sec:results}

We present results of electron scattering on $^4$He for two kinematics, ${\bf q}=300, 400$ MeV/c for which the Rosenbluth separation has been performed. In this regime the LIT-CC results using a chiral Hamiltonian are fully converged. At the same time, the momentum is high enough to justify the impulse approximation in the SF approach.

\begin{figure}[h]
\centering
\includegraphics[width=0.48\textwidth]{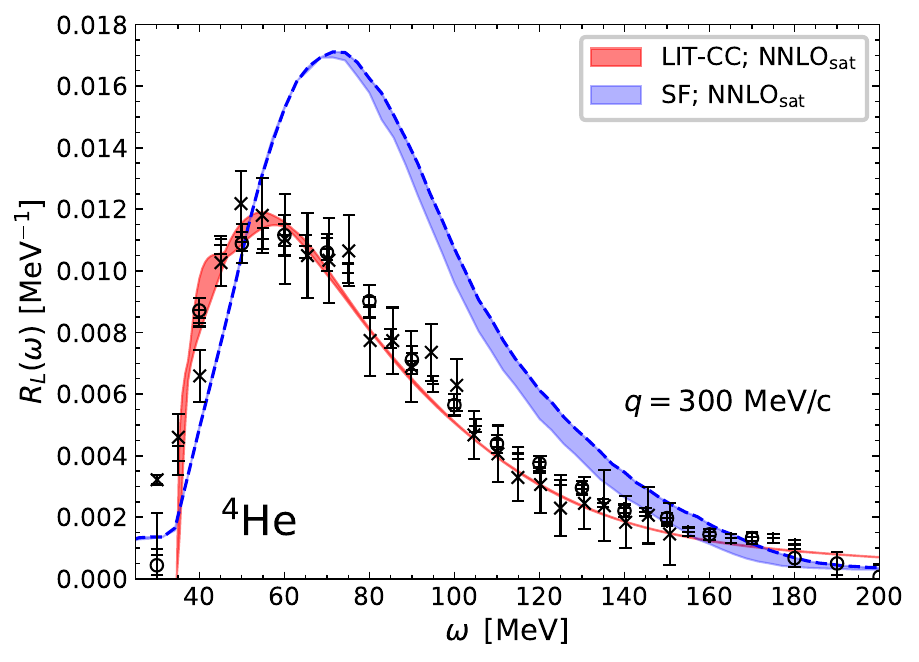}
\includegraphics[width=0.48\textwidth]{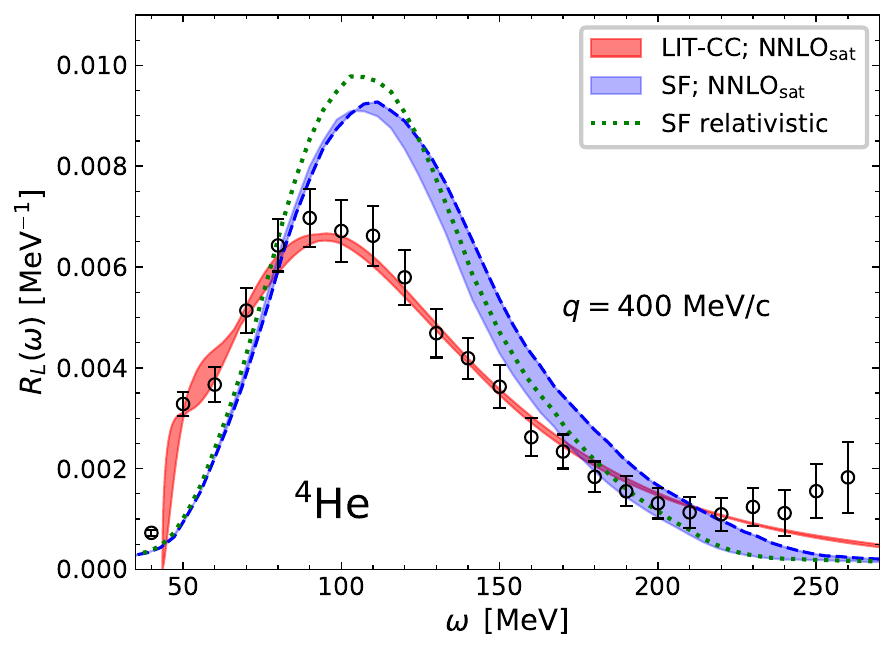}
\caption{Longitudinal response for two values of momentum transfer ${\bf q}=300, 400$ MeV/c. The uncertainty of LIT-CC method (red band) comes from the inversion procedure. The blue dashed line corresponds to the nonrelativistic SF calculation, while the blue band shows the difference from the relativistic case. The dotted green curve shows the fully relativistic calculation from Eq.~\eqref{eq:IArel}. Data taken from Refs.~\cite{Zghiche:1993xg,Carlson:2001mp,Dytman:1988fi}.}
\label{fig1}
\end{figure}

\begin{figure}[h]
\centering
\includegraphics[width=0.48\textwidth]{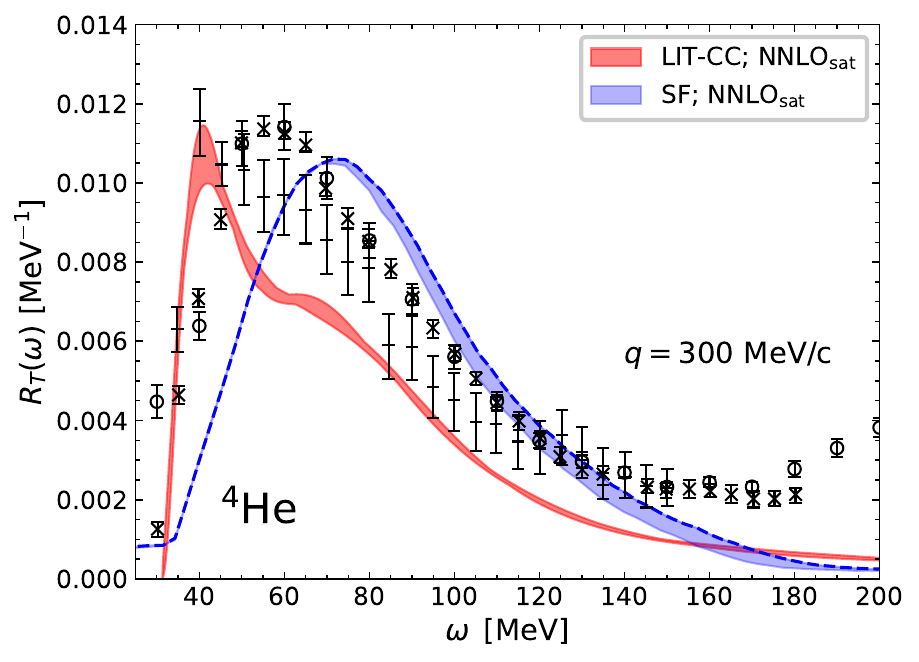}
\includegraphics[width=0.48\textwidth]{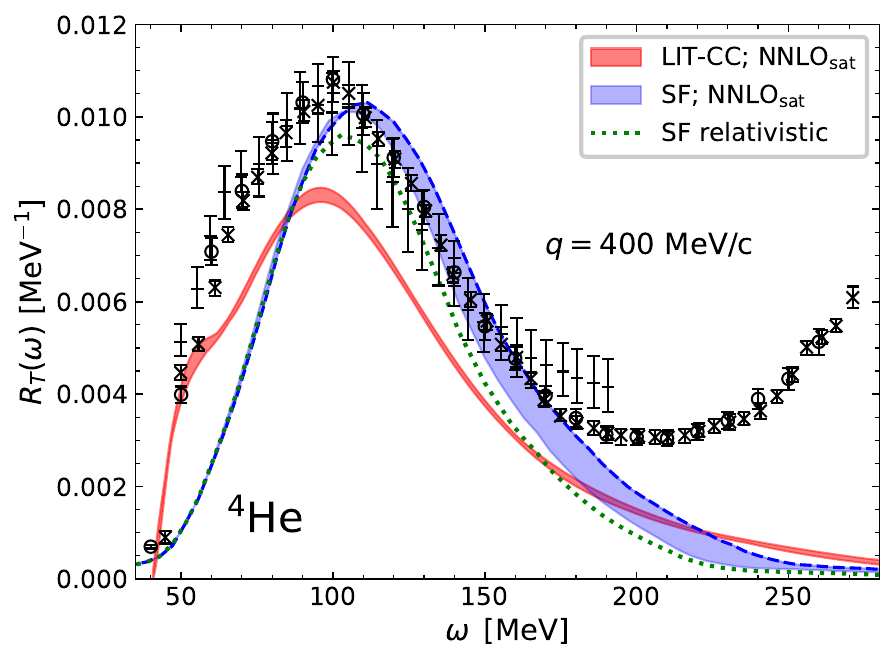}
\caption{Similar to Fig.~\ref{fig1} but for the transverse response.}
\label{fig2}
\end{figure}

Disentangling the longitudinal and transverse responses allows for a more thorough analysis, in which the final state interactions, relativistic effects and the role of two-body currents can be addressed. It is well known that the two-body contribution to the charge operator appears at a higher chiral order, making the longitudinal response a good starting point for the analysis. In fact, in Fig.~\ref{fig1}, we show that the LIT-CC results, which consistently treat final state effects, lead to a  very good agreement with the data already at the level of one-body operator.
The difference between the LIT-CC result and SF (dashed blue line) can be fully prescribed to final state interactions.
This comparison reveals an advantage of LIT-CC in resolving the fine details of nuclear excitations at low energy transfers. In this region (below the QE peak), the SF underestimates the strength since the dynamics of the system is more complicated than a single-nucleon knock-out process.
Within the SF approximation the response is smooth, peaking approximately at $\omega\approx{\bf q}^2/2m$, which corresponds to the kinematics of a process in which a free nucleon receives the whole momentum transfer $q$ and the final knocked-out nucleon is assumed to be on-shell. Also, all interactions with the residual system are neglected. This approximation leads to an enhancement of the QE peak (around $30\%$ at this kinematical setup) and a shift towards higher energy transfers by around 15 MeV. This behaviour can be understood in the mean-field picture: we neglect the potential energy of the outgoing nucleon which would change its dispersion relations in the nuclear medium.
Although it is not straightforward to assess the magnitude of relativistic effects within the LIT-CC, one can account for them within the SF formalism. In Figs.~\ref{fig1}, \ref{fig2} the blue band shows the difference between the nonrelativistic and relativistic kinematics (see Eqs.~\eqref{eq:IAnonrel} and \eqref{eq:IAsemirel}). The dashed blue line corresponds to the nonrelativistic kinematics. While for ${\bf q}=300$ MeV/c momentum transfer the effect is small, it becomes visible for ${\bf q}=400$ MeV/c. It  affects the  high energy tail of the response, above the QE peak. Since this is a purely kinematical effect, we would expect that its magnitude is similar for  the LIT-CC method.

The transverse response, see Fig.~\ref{fig2}, shows a somehow different behaviour. The LIT-CC results underestimate the experimental data by around $20\%$. This can be expected, since for the electromagnetic current operator the two-body currents appear at lower chiral order and were shown to produce an enhancement in several few-body calculations~\cite{Lovato:2015qka,Lovato_2016}. 
At ${\bf q}=300$ MeV/c we also observe a low-lying structure which originates from the magnetic part of the transverse current.
With respect to the LIT-CC, the SF predicts more strength and shifts the QE peak towards higher energy transfers. At the same time, it leads to a reasonable agreement with the experimental strength (although still shifted by 15 MeV). 
For ${\bf q}=400$ MeV/c momentum transfer we also show with the green dotted line in Figs.~\ref{fig1}, \ref{fig2} the results of a fully relativistic calculation, Eq.~\eqref{eq:IArel} (both in kinematics and one-body currents). The difference is of the order of a few percent in the strength of the response.
Finally, at higher energies, above $200$ MeV, the experimental data reveals additional mechanisms above the QE peak. Pion production, which is not accounted for in our method, starts playing an important role and dominates the transverse response. 

It is important to point out that the analysis of nuclear responses give insights which could be easily missed when considering the inclusive cross section only. In fact, for the kinematics in which $R_T$ dominates (backward scattering angles), the SF approach could predict a result close to experiment since the final-state-interactions effects would compensate for the lack of two-body currents.

The analysis also reveals that the relativistic effects of the LIT-CC should be already important for ${\bf q}=400$ MeV/c and they would affect mostly the distribution tail. One could account for them following a procedure of Ref.~\cite{Efros:2005uk}, i.e. performing calculation in a boosted reference frame in which the relativistic effects become smaller.

\section{Conclusions}

We conducted a comparison between two approaches for calculating the lepton-nucleus cross section to investigate the interplay of various effects: final state interactions, two-body currents, and relativistic effects. Specifically, we analyzed the longitudinal and transverse responses for electron-$^4$He scattering, utilizing available Rosenbluth separated data in the intermediate momentum transfer regime. Both the LIT and SF calculations were derived within the coupled-cluster theory using a chiral nuclear Hamiltonian. This shared theoretical framework enabled a meaningful comparison.

The LIT-CC method represents the current state-of-the-art approach for calculating nuclear responses while consistently considering final state interactions. However, it is limited to the nonrelativistic regime. Conversely, the SF formalism, which can extend to the relativistic energies, does not incorporate final state interactions.

We point out that the lack of two-body currents, crucial for the transverse response, might lead to a good comparison of the SF method only with one-body currents. 
Moreover, the relativistic effects play a non-negligible role already at ${\bf q}=400$ MeV/c and can lead to a suppression of the distribution tail, beyond the QE peak.

The conclusions drawn from our analysis of electron scattering processes can be directly applied for the neutrino induced reactions, thereby making a direct impact on the theoretical studies for neutrino oscillation experiments.

\bmhead{Acknowledgements}
This project has received funding from the European Union’s Horizon 2020 research and innovation programme under the Marie Skłodowska-Curie grant agreement No.~101026014.
This work was supported by the Deutsche
Forschungsgemeinschaft (DFG) 
through the Cluster of Excellence ``Precision Physics, Fundamental
Interactions, and Structure of Matter" (PRISMA$^+$ EXC 2118/1) funded by the
DFG within the German Excellence Strategy (Project ID 39083149). Computer time was provided by the supercomputer MogonII at Johannes Gutenberg-Universit\"{a}t Mainz.

\bibliography{sn-bibliography}

\end{document}